\documentclass{jfm}

\usepackage{natbib}
\usepackage{latexsym}
\usepackage{amsmath}
\usepackage{upmath}
\usepackage{amsbsy}
\usepackage{amssymb}
\usepackage{amsgen}
\usepackage{amsfonts}
\usepackage{hyperref}
\usepackage{graphicx}
\usepackage{ulem}
\usepackage[english]{babel}
\usepackage[utf8]{inputenc}
\usepackage{lmodern}
\usepackage[T1]{fontenc}

\title{Liquid jet eruption from hollow relaxation}
\author[É.~Ghabache, T.~Séon and A.~Antkowiak] 
{É\ls L\ls I\ls S\ls A\ls B\ls E\ls T\ls H\ns G\ls H\ls A\ls B\ls A\ls C\ls H\ls E,
\ns T\ls H\ls O\ls M\ls A\ls S\ns  S\ls É\ls O\ls N\ns\and
\ns A\ls R\ls N\ls A\ls U\ls D\ns A\ls N\ls T\ls K\ls O\ls W\ls I\ls A\ls K\ns
 }
\affiliation{Sorbonne Universités, UPMC Univ Paris 06, CNRS, UMR 7190 Institut Jean Le Rond d'Alembert, F-75005 Paris, France.}

\begin{document}
\maketitle

\begin{abstract}
A cavity hollowed out on a free liquid surface is relaxing, forming an intense liquid jet. Using a model experiment where a short air pulse sculpts an initial large crater, we depict the different stages in the gravitational cavity collapse and in the jet formation. Prior eversion, all cavity profiles are found to exhibit a shape similarity. Following hollow relaxation, a universal scaling law establishing an unexpected relation between the jet eruption velocity, the initial cavity geometry and the liquid viscosity is evidenced experimentally. On further analysing the jet forms we demonstrate that the stretched liquid jet  also presents shape similarity.
Considering that the jet shape is a signature of the initial flow focusing, we elaborate a simple model capturing the key features of the erupting jet velocity scaling.
\end{abstract}

% _________________________________________
%
%   I :   I N T R O D U C T I O N
% _________________________________________
%

\section{Introduction}

%%%%%%%%%%%%%%%%%%%% DISCUSSION %%%%%%%%%%%%%%%%%%%%%%%%%%%%%%%%%%%%%%%%

% Dans la dissipation dire que elle augmente qd nu augmente et L diminue 
% Discussion sur L qui scale à la fois avec la zone de redirectin de vitesse qui si elle se ressert augmente la vitesse et sur la zone de dissipation qui si elle diminue augmente la dissipation et dinminue la vitesse en augmentant le cisaillement.
% La viscoisté commence a jouer a partir de 400 avant cest "inviscide"
% Chez Lorenceau dz/dt diminue qd Z_0 augmente moi je dis que qd Z_0 augmente la zone de dissipation augmente et la dissipiation diminue et la vitesse augmente.

%%%%%%%%%%%%%%%%%%%%%%%%%%%%%%%%%%%%%%%%%%%%%%%%%%%%%%%%%%%%%%%%%%%

Vertical-take-off-and-landing craft can experience loss of visibility or damage when landing over water or a soft terrain, such as mud or sand, as a consequence of the soil erosion by a gaseous jet \citep{Barton1968}.
In the steel industry, the basic oxygen conversion process utilizes a supersonic jet of oxygen impinging on molten iron to convert it into steel. 
%Both these examples involve the steady deformation of a free surface imparted by a gaseous jet. 
The generic occurrence of such gaseous jets shaping steady cavities has sparked a number of studies involving model experiments for the last 50 years \citep{Banks1963, Cheslak1969}. 

Violent free surface deformation can also be observed when an object impacts a liquid surface and induces an unsteady cavity whose collapse often exhibits an intense liquid jet \citep{Worthington1883}. This phenomenon has been extensively studied, from the initial stages of contact \citep{Korobkin1988}, through the dynamics of transient cavities produced thereof (see \textit{e.g.} \citealp{Bisighini2010} for drop impact or \citealp{Birkhoff1957, Duclaux2007} for solid body impact) up to jet formation \citep{Hogrefe1998, Gekle2009}.
%, where jetting is fed by the local flow around the base of the jet.
Such a jet is a classic signature of hollow crater-like relaxation, and appears at scales ranging from champagne bubbles \citep{Liger-Belair2009} to geological craters central peak \citep{Melosh1989}.

\begin{figure}
\noindent\includegraphics[width=\hsize]{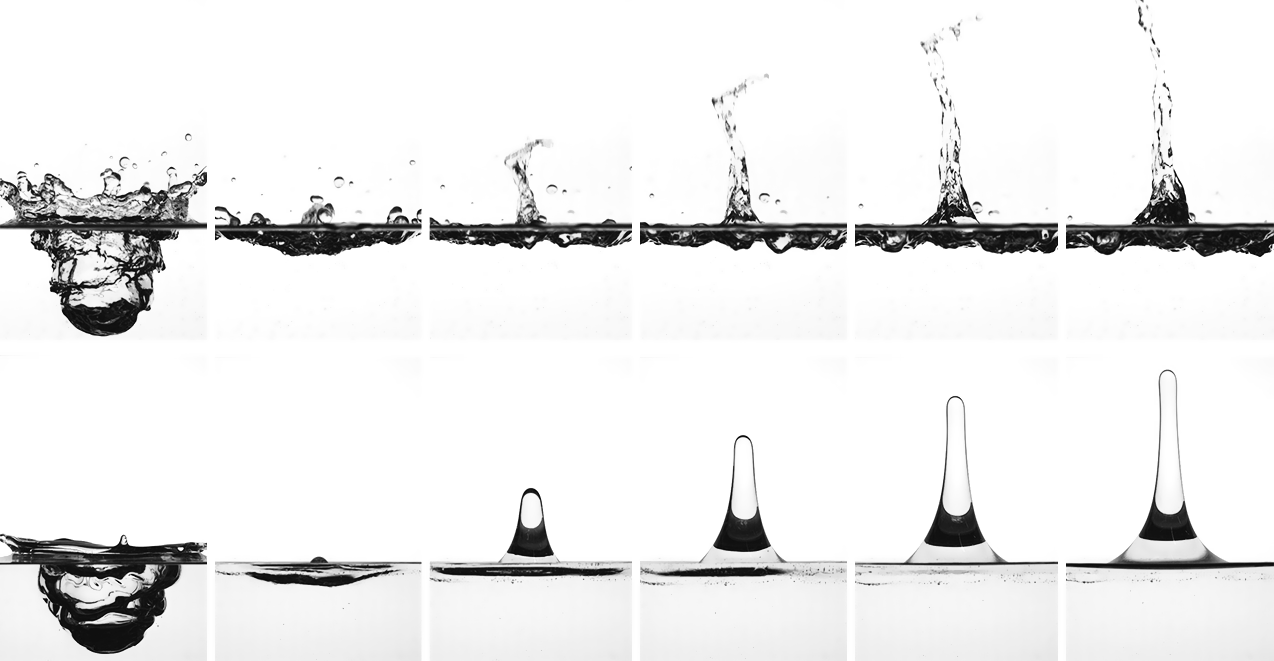}
\caption{Time sequences of the jet induced by the relaxation of a centimeter-sized hollow at the free surface of two different liquids : water ($\mu=1$ mPa.s) on the top and a water-glycerol mixture of viscosity $\mu=833$ mPa.s on the bottom. Each sequence starts with a snapshot of the cavity, shaped by an air pulse, at its maximum depth $H$ and about to relax. On the second image the cavity is reversing giving rise to a high speed jet observed on the following images. For the top sequence in water the time lapse between the two first snapshots is $\Delta t_{\text{1-2}}$ = 36~ms and between the following images is $\Delta t$ = 7.5 ms. The maximum hollow depth is $H$ = 4.6~cm, the width is $L$ =~6.3 cm and the jet velocity is V$_{\text{tip}}$= 3.59 m.s$^{-1}$. For the bottom sequence in the viscous solution : $\Delta t_{\text{1-2}}$ = 60 ms, $\Delta t$ = 27.5 ms, $H$ = 6 cm, $L$ = 10 cm and V$_{\text{tip}}$= 1.85~m.s$^{-1}$.}
\label{fig:sequence_grav}
\end{figure}

In this paper, we propose to investigate the characteristics of the erupting jet  following large hollow relaxation in relation with the initial cavity geometry. We consider a model experiment where the crater, shaped using a short air pulse, relaxes sparking a fast inertial liquid jet (see Fig.~\ref{fig:sequence_grav}). 
After a description of the important stages in the jet development, we explore the dependence of the erupting jet velocity with the liquid properties and the cavity geometry. As a result we propose a universal scaling law for the jet velocity. We further demonstrate that both the initial cavities and the outcoming jet exhibit shape similarity. We build up on this observation and finally propose a simple model capturing the key features of the jet velocity dependence with the cavity shape.

\section{Experimental description}

\subsection{Experimental set-up}

Our experiment consists in shaping a hollow at a free liquid surface by blowing a short air pulse from a nozzle: as soon as the airflow stops, the cavity relaxes under gravity to form a liquid jet. Both the depth $H$ and the width $L$ of the initial cavity can be tuned by adjusting properly the air pulse intensity and spreading: the higher the air pulse velocity, the deeper the cavity and the larger the height of the nozzle from the free surface, the wider the cavity. Typical heights and widths range from 3 cm to 10~cm and typical cavity aspect ratios $H/L$ from 0.25 to 1.1 (a hemispherical cavity would have an aspect ratio $H/L$ = 0.5). The liquid, contained in a transparent tank (30 cm $\times$ 30~cm $\times$ 35 cm), is a water-glycerol mixture. 
Depending on the proportions of water and glycerol, the liquid viscosity can be set in the range $\mu=$1~mPa.s -- 1300 mPa.s. Note that temperature-induced variations have been accounted for throughout this study. 
The mixture density ranges from $\rho~=~$1000~kg.m$^{-3}$ to 1250 kg.m$^{-3}$.
%, and the surface tension of the mixture ranges from 72 to 64 mN.m$^{-1}$. 
The hollow relaxation and jet dynamics are analyzed through ultrafast imaging. Typically, images of the back lit scene are obtained at 2000 frames per second using a digital high-speed camera (Photron SA-5).

\begin{figure}
\begin{center}
\noindent\includegraphics[width=12cm]{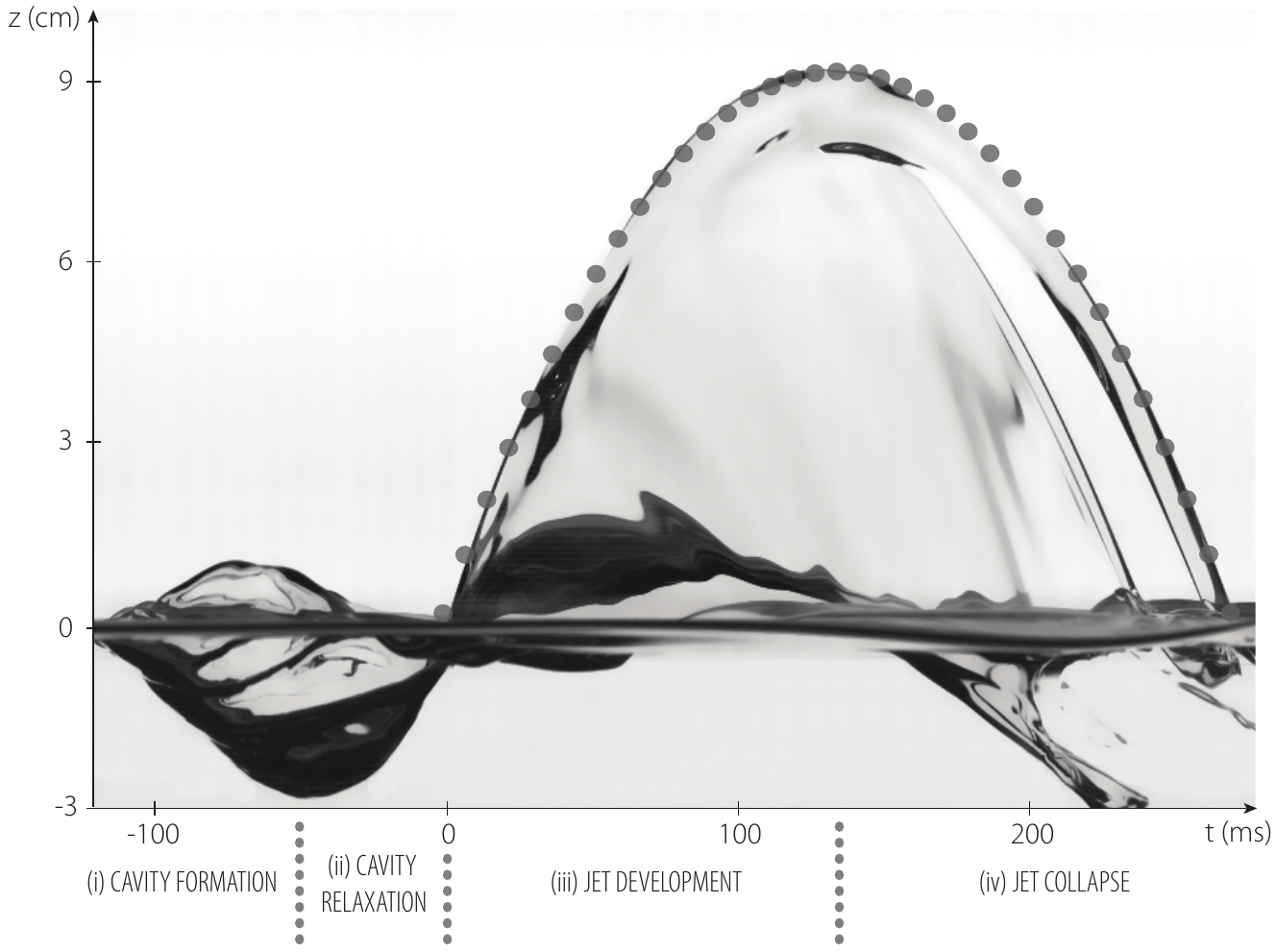}
\caption{Spatio-temporal diagram presenting \textit{(i)} the cavity formation, \textit{(ii)} the cavity relaxation and reversal, \textit{(iii)} the jet development and \textit{(iv)} the jet collapse. These four steps are indicated in the timeline. The time origin ($t = 0$) is taken when the jet crosses the free surface $z = 0$, and the local slope gives the jet velocity V$_\text{tip}$. % The snapshot in the negative times illustrate the maximum depth cavity separating (i) and (ii), the one in the positive time illustrate the maximum hight jet separating (iii) and (iv). 
The dotted curve represents the free-fall parabola corresponding to this jet eruption velocity V$_{\text{tip}}$.
In this particular case the liquid viscosity is $\mu=10$~mPa.s, the jet velocity is V$_{\text{tip}}=1.33$~m.s$^{-1}$, the cavity maximum depth and width are $H$=2.9~cm and $L$=5.5~cm.} 
\label{fig:spatio}
\end{center}
\end{figure}

\subsection{Qualitative overview of the jet dynamics}

Figure \ref{fig:sequence_grav} illustrates two typical jetting events induced by the relaxation of a large surface hollow (whose width $L$ is greater than the capillary length). The top sequence has been obtained with pure water as a working liquid while the 
bottom one corresponds to a water-glycerol mixture eight hundred times more viscous. On the first images, the depression hollowed 
out in the liquid by the air pulse has reached its maximum depth~$H$ and is about to relax. Just after maximal expansion, the liquid 
starts flowing in so as to fill the entire cavity. But this filling process is not just a simple relaxation to the minimal potential energy state 
(\textit{i.e.} flat surface): during the collapse the cavity curvature reverses, prelude to the eruption of a violent liquid jet \citep{Lavrentiev1980}. It is remarkable that the developing vertical jet can reach heights larger than ten times the initial depth $H$ of the cavity. This is a strong indication that the pressure built up during the collapse far exceeds the hydrostatic pressure: the gravity-driven flow focuses energy and momentum. Such focusing is generic in jet eruption and can be observed in situations as diverse as impact-driven \citep{Antkowiak2007}, needle less injection systems \citep{Tagawa2012} or even granular jets \citep{Lohse2004}. 

\noindent Insights into the role played by viscosity can readily be gained by comparing the two sequences of Fig.~\ref{fig:sequence_grav}. 
Apart from its obvious smoothing action on the cavity, jet and lip surfaces, viscosity also proves to slow down the jet, as the time between each snapshot of Fig.~\ref{fig:sequence_grav} is much longer for the viscous case ($ \Delta 
t_{\text{viscous}}=27.5$ ms > $\Delta t_{\text{water}}=7.5$ ms).  Noteworthy enough, this slowing action of viscosity is achieved \textit{before}  jet eruption as evidenced by the jet velocity seen to be  constant in both sequences, at least before the final deceleration due to gravity. 

%Conversely, the jet in the viscous liquid is clean and its profile 
%easy to extract quantitatively.
 % dire que le vitesse est constante sur les 4 premieres images en eau et sur les 3 premieres en visqueux

\noindent A global overview of the jetting event is provided Fig.~\ref{fig:spatio} with a spatio-temporal diagram depicting the whole jet dynamics with the important stages in its chronology: \textit{(i)} cavity formation up to maximum depth~$H$, \textit{(ii)} cavity relaxation and reversal, \textit{(iii)} jet eruption and development and \textit{(iv)} final jet collapse and/or disruption. Right after eruption, the jet velocity V$_{\text{tip}}$, measured on the spatio-temporal diagram, is seen to be important  with respect to the characteristic gravito-inertial velocity $\sqrt{gH}$  (here V$_{\text{tip}}=1.33$~m.s$^{-1} > \sqrt{gH}$=0.53~m.s$^{-1}$). Actually, in all the reported experiments, the Froude number Fr=V$_{\text{tip}}/\sqrt{gH}$ was systematically observed to be greater than unity. Once shooted out above the free surface, the jet tip follows a ballistic trajectory characterized by a free-fall parabola (see dotted curve Fig.~\ref{fig:spatio}), even for liquids a hundred times more viscous than water (see Supplemental Material). This ballistic behaviour demonstrates that viscosity or surface tension barely affect the jet development. The purely inertial description assumption for the jet dynamics employed in this paper results from this key observation.% C'est un peu chaud de parler du froud maintenant non ? (Arnaud : non c'est bon)
%($\mu$ = 833 mPa.s, H = 4.9 cm and L = 9.6 cm)

%In the following we investigate how the jet eruption velocity V$_\text{tip}$, completely determining the ballistic jet evolution, depends on the initial cavity geometry and on the liquid properties.

% _________________________________________
%
%   I I I :   U N I V E R S A L   S C A L I N G   F O R   T H E   J E T
%
%                      E R U P T I O N   V E L O C I T Y  
% _________________________________________
%

\section{Universal scaling for the jet eruption velocity}	

\begin{figure}%[htd]
\begin{center}
\noindent\includegraphics[width=12cm]{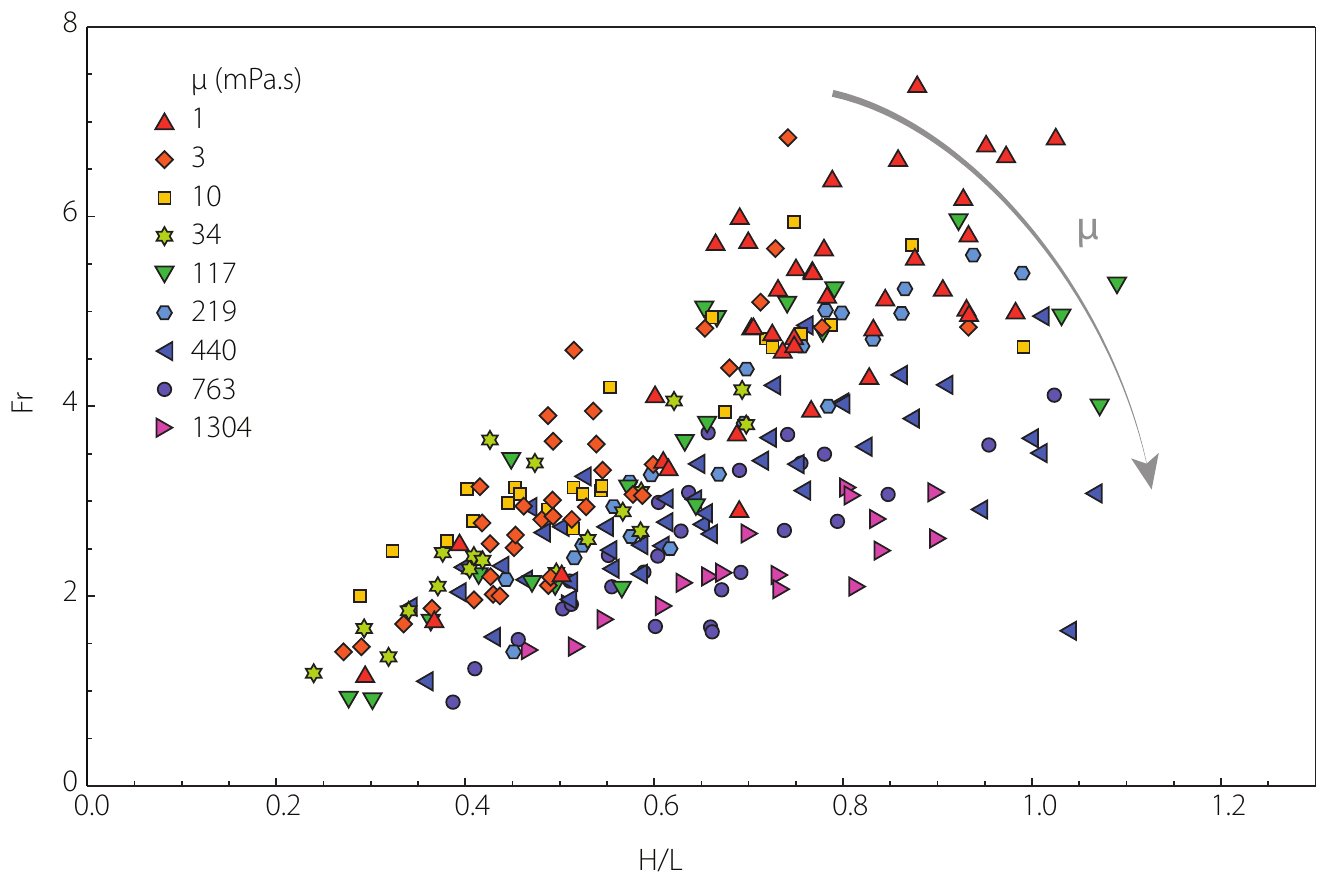}
\caption{The Froude number Fr=V$_{\text{tip}}/\sqrt{gH}$ increases with the aspect ratio $H/L$. The damping action of viscosity $\mu$ naturally lowers the eruption velocity (see arrow).} 
\label{fig:froude}
\end{center}
\end{figure}

We now investigate how the jet eruption velocity V$_\text{tip}$, completely determining the ballistic jet evolution, depends on the initial cavity geometry and on the liquid properties.
%To look at the jet dynamics dependence on the control parameters, the jet tip velocity V$_{\text{tip}}$ is measured for jets induced by various aspect ratios $H/L$ cavity and made in solutions with different viscosities. 
%
We start with nondimensionalising the eruption velocity V$_{\text{tip}}$ with the natural gravito-inertial velocity scale $\sqrt{gH}$ and the depth $H$ with the width $L$. The resulting relation between the Froude number Fr=V$_{\text{tip}}/\sqrt{gH}$ and the cavity aspect ratio $H/L$ is represented Fig.~\ref{fig:froude} for nine mixtures with viscosities distributed in the range $\mu=$1 mPa.s to 1300 mPa.s. % c est tres disperse car ... ?
Two experimental facts directly emerge from the picture: first the eruption Froude number builds up with increasing cavity aspect ratio (\textit{i.e.} with narrower cavities) and second, the viscosity has a damping action on the jet velocity. While the latter observation is certainly to be expected, the dependence of the eruption Froude number with $H/L$ might be surprising. Indeed, while studying liquid invasion in initially empty pipes, \citet{Lorenceau2002} reported that the liquid column rising velocity was systematically found to be very close to $\sqrt{gH}$. In this  `rigid cavity filling' experiment, the Froude number therefore tends to 1 irrespective of the tube radius. The significant difference observed in velocity scaling in otherwise analogous experiments will later prove to be related to the energy and momentum focusing occurring in cavity relaxation.
%
%tube studying an analogous setup of liquid invasion of observed that the velocity of the rising liquid Two clear trends appear from this figure:  the Froude number increases with aspect ratio and decreases with viscosity. The increase with $H/L$ might be surprising, indeed, from such a system exhibiting a balance between buoyancy and inertia, we would expect a velocity proportional to the gravito-inertial velocity V$_{\text{tip}} \propto \sqrt{gH}$ as it appears 
%CAVITY FILLING
%for instance in large bubble dynamics \cite{Davies1950} or inertial gravity current \cite{Benjamin1968}. 
% liquid column partially immersed in a bath of liquid
%or for a rising liquid column in an immerged pipe \cite{Lorenceau2002}.

\begin{figure}%[htd]
\begin{center}
\noindent\includegraphics[width=12cm]{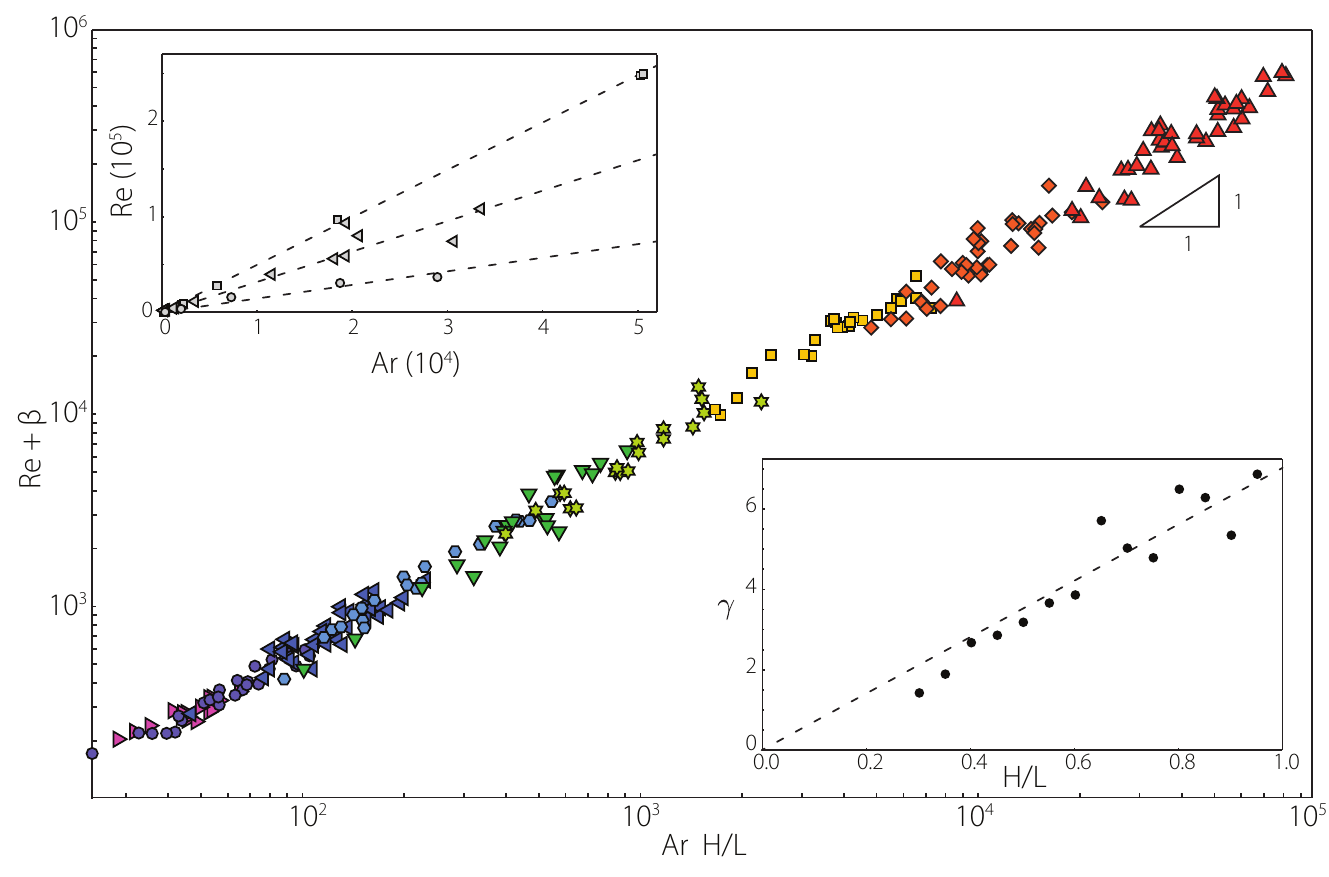}
\caption{Experimental relation between $\text{Re}+\beta$ and $\text{Ar}\frac{H}{L}$ for the nine sets of experiments represented Fig.~\ref{fig:froude}. Top left inset: Reynolds number Re versus Archimedes number Ar for subsets of experiments having constant aspect ratio $H/L =0.3$, $0.5$, $0.7$. Each set of data is organized along a line of slope $\gamma(H/L)$ and all have the same $y$-intercept $\beta$. Bottom right inset: Linear dependence of the slope $\gamma(H/L)$ with the cavity aspect ratio $H/L$: $\gamma(H/L) = \alpha \frac{H}{L}$. The constants $\alpha$ and $\beta$ defining the master curve are given by $\alpha = 6.59$ and $\beta= 104$.} 
\label{fig:scaling_law}
\end{center}
\end{figure}
%Dimensional analysis : 
\noindent Intending to obtain a universal scaling law for the eruption velocity also accounting for the action of viscosity, we identify so far five geometrical or dynamical variables ruling the value of the jet tip velocity:
\begin{equation}
\mathrm V_{\text{tip}} = F\left(\rho, \sqrt{gH}, \mu, H, L\right).
\label{eq:tipvelocity_dependence}
\end{equation}
%the action of viscosityThis problem is entirely characterized by five variables: $H$, $L$, $\rho$, $\mu$ and the gravito-inertial velocity $\sqrt{gH}$. 
\noindent Choosing $L$, $\rho$ and $\mu$ as independant scales, we use dimensional arguments \citep{barenblatt} to recast the former equation into a relation between the dimensionless tip velocity $\text{Re}=\rho $V$_{\text{tip}} L/\mu$, the Archimedes number $\text{Ar}=\rho \sqrt{gH} L/\mu$ and the aspect ratio $H/L$:
\begin{equation}
\text{Re}=\mathcal F\left(\text{Ar},\frac{H}{L}\right).
\label{eq:pi_theorem}
\end{equation}
We investigate experimentally the dependence of Re with Ar by selecting subsets of experiments having constant cavity aspect ratio~$H/L$. The relation between Reynolds and Archimedes numbers is illustrated in the top left inset of Fig.~\ref{fig:scaling_law} for three different~$H/L$. It is instructive to note that each data subset is organized along a  line of slope~$\gamma$. While this slope appears to depend on $H/L$, detailed analysis of the data show that each line presents the same $y$-intercept~$\beta$, regardless of the aspect ratio value. As a result we may rewrite relation~(\ref{eq:pi_theorem}) as $\text{Re}=\gamma(\frac{H}{L}) \text{Ar} -\beta$. Further progress can be obtained by extracting the slope $\gamma(H/L)$ for a number of experimental subsets having various $H/L$, see bottom right inset of Fig.~\ref{fig:scaling_law}. Experimental data clearly demonstrate that the relation between $\gamma$ and $H/L$ is linear: $\gamma=\alpha \frac{H}{L}$, with $\alpha$ a non dimensional constant. This leads to the reformulation of~(\ref{eq:pi_theorem}) as the following scaling law:
\begin{equation}
\text{Re}=\alpha \frac{H}{L} \text{Ar}  - \beta.
\label{eq:ReArHL}
\end{equation}

\noindent Figure~\ref{fig:scaling_law} illustrates the experimental relation between $\text{Re} + \beta$ and $\text{Ar} \frac{H}{L}$ for the nine series of experiments conducted at various initial cavity geometries and different liquid properties. The collapse of all the experimental data on a single master curve of slope unity is remarkable and ranges over almost four decades, thereby confirming the relation between the jet eruption velocity and the initial hollow depth~(\ref{eq:ReArHL}). The dimensional counterpart of this simple and universal scaling law can be expressed as :
\begin{equation}
\text{V}_{\text{tip}} =\alpha\sqrt{gH}\frac{H}{L} - \beta \frac{\nu}{L},
\label{eq:dimensional_scaling_law}
\end{equation}
where $\nu=\mu/\rho$ denotes the liquid kinematic viscosity. Equation~(\ref{eq:dimensional_scaling_law}) makes it clear that the jet dynamics is merely a signature of the gravitational cavity collapse, but also that the offset $\beta$ in relation~(\ref{eq:ReArHL}) has a viscous origin.

\noindent It is interesting to note that, unless experiments are performed with constant aspect ratio cavities, the tip velocity depends not only on the natural gravito-inertial scale $\sqrt{g H}$ but also on the global cavity geometry ($H/L$). This is at variance with other systems involving a balance or a conversion of gravity potential energy into inertia, such as spherical cap bubbles, gravity currents or rigid cavity filling \citep{Lorenceau2002}, where a local measure (typically of the front curvature radius or of the depth) allows to infer the velocity. In other words, a sequence of cavities having the same lateral extent $L$ but different depths $H$ will induce jets with velocities varying with $H^{3/2}$, as opposed to the previous examples where the dependence with $H^{1/2}$ prevails. We relate the apparition of this unexpected exponent to an energy focusing phenomenon, as we show in the next section. 

\noindent Though of inviscid nature, (\ref{eq:dimensional_scaling_law}) indicates that the jet nonetheless experiences a neat braking from viscous processes. % before eruption (as mentioned in \S 2). 
As mentioned earlier (\S 2), in our experiment viscous damping occurs for the most part before jet eruption. 
We therefore expect dissipative phenomena to take place underneath the jet along with the strong deformation associated with the conversion of radial inflow into axial motion.
This interpretation sheds light on the reason why viscosity does not appear in a prefactor but rather in an offset. 
Such a functional dependence for the dissipative action was already remarked in \citet{Tagawa2012}; there, the energy loss was associated to the heating of liquid by a laser beam. 

\noindent It is eventually worth noting that in the quite different context of jetting event sparked off during large bubble formation, \citet{Seon2012} did observe both the $H^{3/2}$ dependence of the jet velocity ($H$ being a measure of the bubble height) and an analogous viscous offset, therefore
putting in perspective the universality of the proposed scaling law.

\section{Shapes and dynamics:\\ \quad\, Experimental rationalization and simple model}

%In order to grasp the eruption velocity scaling law, we present in the following a detailed description of the cavity and jet kinematics. This understanding will allow us to derive a simple energy balance model capturing the dependence of the jet velocity with the cavity geometry.

\noindent 
In the following we present a detailed description of the cavity and jet kinematics. In addition to improve our understanding of the cavity expansion and the jet dynamics, a knowledge of the shapes -- that are a signature of the cavity relaxation and the related flow focusing -- will allow us to derive a simple energy balance model capturing the dependence of the jet velocity with the cavity geometry.

%We present in the following a detailed description of the cavity and jet kinematics. This understanding will allow us to derive a simple energy balance model capturing the eruption velocity scaling law.
%In order to grasp the dependence of the jet eruption velocity with the cavity geometry,

\subsection{Cavity shape similarity} 

%The cavity is created at the free surface by an air jet impinging on a liquid for a short time. 
Numerous experimental studies have examined the deformation imparted to a free surface by an impinging gas jet, starting with the work of \citeauthor{Banks1963} in the early sixties \citep{Banks1963, Cheslak1969}. These studies have described how the geometrical properties of the cavity are linked to the blowing conditions (gas velocity, nozzle height and diameter). In particular, most of these works have considered steady cavity profiles to be decently fitted with a parabola. Conversely, unsteady liquid cratering dynamics have been investigated mainly in the case of point source, obtained for example by means of drop impact \citep{Bisighini2010}. In this type of experiment, the cavity shape is very close to an hemisphere and the dynamics of expansion-retraction is well described by Rayleigh-Plesset equation. In our experiment, the pressure forces shaping the cavity are distributed along the cavity, \textit{i.e.} not localized.

\begin{figure}%[ht]
\begin{center}
\noindent\includegraphics[width=\hsize]{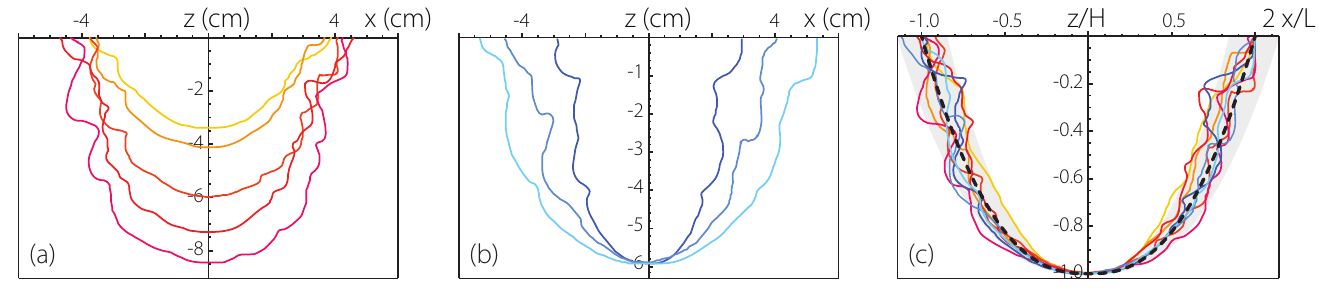}
\caption{Different cavity profiles with (a) almost the same width ($L$) and different heights ($H$) and a liquid viscosity of $\mu=219$~mPa.s, and (b) the same height, different widths and $\mu=440$~mPa.s. (c) The same cavity profiles after rescaling: $z/H$ vs $2x/L$.  % the horizontal coordinate is divided by $L$ and the vertical coordinate by $H$
They all gather in the gray region: $z=x^{2.5\pm 0.1}-1$ exhibiting a similar cavity shape.The black dashed line represent the curve $y=x^{2.5}-1$.} 
\label{fig:cavity}
\end{center}
\end{figure}

\noindent Figure~\ref{fig:cavity} reports cavity profiles at their deepest -- that is, right before relaxation -- for eight different experiments corresponding to various values of the gas velocity (Fig.~\ref{fig:cavity}a) and of the nozzle height (Fig.~\ref{fig:cavity}b). We observe that by increasing the velocity of the air jet, deeper cavities can be dug keeping their width $L$ almost constant. Conversely moving up the nozzle allows for wider cavities to be formed. This enables us to build up initial hollows with various dimensions and aspect ratios. It is remarkable that all these different cavities exhibit shape similarity. This is demonstrated Fig.~\ref{fig:cavity}c where the eight cavity profiles are plotted on the same graph after rescaling the horizontal coordinates with the cavity radius $L/2$ and  the vertical coordinates with the cavity depth $H$. All cavity contours gather onto a universal profile. The best fit for this master curve is given by  $z/H=( 2 x/L )^{2.5\pm 0.1}-1$, with $z$ and $x$ the vertical and radial coordinates (black curve in Fig.~\ref{fig:cavity}). The shaded area represents the error bar taken on the exponent.

% dans la biblio voir si on trouve que les profils sont autosimilaires, et comment ils voient qu'ils sont parabolique
% Dire aussi que la vitesse n'est jamais totalement nulle dans le liquide

\subsection{Jet dynamics and shape similarity   }

We now turn to the kinematics of jet formation, that starts with the cavity eversion. Figure~\ref{fig:cavity}a shows a typical sequence of successive jet profiles ($\Delta t=$ 5 ms between each) extracted from the same experiment. Here $z=0$ represents the far free surface. The roughly constant spacing between the successive jet tips is again an indication for the constant initial jet velocity (high Froude number eruption), consistent with the spatio-temporal diagram analysis (Fig.~\ref{fig:spatio}). The transverse time dependence of the jet is assessed by representing the evolution of the jet section $r_{\text{jet}}^2(z,t)$ (disregarding the factor $\pi$) at five positions on the vertical axis $z$ (Fig.~\ref{fig:cavity}b). The linear trend curves  capture satisfyingly the evolution of the jet section for every heights, revealing that the jet radius varies as the square root of time. The experimental data therefore suggest the following functional form for $r_{\text{jet}}(z,t)$ :
\begin{equation} \label{eq.r}
	r_{\text{jet}}(z,t)=A(z)\sqrt{t+t_{0}}
\end{equation}	
where the prefactor $A(z)$ is a function of the height $z$ only. % a constant that be different of 0 as we can notice on the graph. %distinguish ?
This dependence of the jet shape with height is showed in the following to be  completely constrained by the dynamics. 

\begin{figure}%[htd]
\begin{center}
\noindent\includegraphics[width=\hsize]{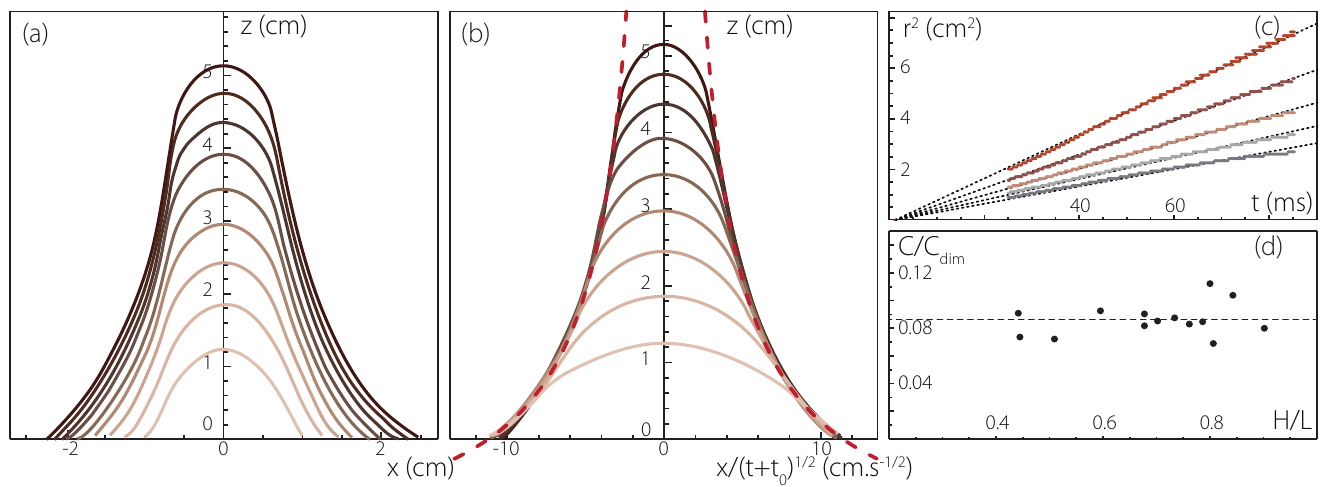}
\caption{(a) Experimental jet profiles evolution with time following a cavity relaxation ($H=4.9$cm, $L=9.6$cm V$_\text{tip}=1.41$m.s$^{-1}$) in a liquid of viscosity $\mu=763$~mPa.$s$. The interval between two profiles is 5ms.  (b) The same jet profiles after rescaling of the $x$ coordinate by $\sqrt{t+t_0}$. The dashed line red curve follows the equation $z=C x^{-1} -z_0$.
(c) Time dependence of the square radius of the jet ($r^2$) at different locations $z$ in the jet. (d) Experimental relation between the jet radius prefactor $C$ non-dimensionalized by $C_\text{dim}~= (gH)^{1/4} L^{3/2}$ and the cavity aspect ratio $H/L$.} 
\label{fig:jet_profile}
\end{center}
\end{figure}

\noindent As mentioned earlier, experimental evidences suggest that the dynamics of the jet is barely affected by viscosity, capillarity or even gravity ($\text{Fr} > 1$), at least for the range of parameters studied. Exploiting both the inertia dominated character of the jet and its slenderness, we can formulate the governing equations expressing mass and momentum conservations in a one-dimensional setting (see \citet{Eggers1997} for an historical account on one-dimensional approximations):
\begin{subequations}
\begin{eqnarray} 
	\frac{\partial r}{\partial t}+u\frac{\partial r}{\partial z} & = & -\frac{1}{2}r\frac{\partial u}{\partial \, z}, \label{eq:mass_conservation} \\[1em]
%\end{equation}
%\begin{equation}
	\frac{\partial u}{\partial t}+u \frac{\partial u}{\partial z} & = & 0,
\label{eq:momentum_conservation}
\end{eqnarray}
\end{subequations}
noting $r(z,t)$ the local jet radius and $u(z,t)$ the local velocity in a fluid slice.

\noindent For non-shock forming feeding conditions, the velocity within the jet (\textit{i.e.} the solution of equation~(\ref{eq:momentum_conservation})) is naturally attracted by the self-similar relaxation wave velocity profile:
\begin{equation}
	u(z,t)=\frac{z+z_0}{t+t_0},
	\label{eq:relaxation_wave}
\end{equation}	
where $z_{0}$ and $t_{0}$ are space and time shifts. At the tip, the velocity field within the jet $u$($z_{\text{tip}}$) must match the tip velocity V$_{\text{tip}}$. This kinematic condition constrains $z_0$ and $t_0$ via the relation V$_{\text{tip}} =  z_{0}/t_{0}$. 

\noindent Analogously we now investigate jet shape similarity, by looking for solutions $r(z,t)$ of the mass conservation equation~(\ref{eq:mass_conservation}) in the form of monomials~$z^a t^b$. A one-parameter family of solutions such that $a =-(2 b+1)/2$ is readily obtained. Using the linear time dependence of the jet section established experimentally, we infer that the particular exponent pair  ($a$, $b$) = ($-1$,$\frac{1}{2}$) is selected:
\begin{equation}
	r_{\text{jet}}(z,t)=C\frac{\sqrt{t+t_{0}}}{z+z_{0}}, 
	\label{eq:radius_dependence} 
\end{equation}
where $C$ is a constant prefactor. 

\noindent This result sheds light on the experimental jet profiles reported Fig.~\ref{fig:jet_profile}a. Representing these profiles as a function of the self-similar variable $r/\sqrt{t+t_{0}}$ reveals that the jet actually develops in a self-similar sheath, see Fig.~\ref{fig:jet_profile}b. It is remarkable that the jet flanks adopt a perfect hyperbolic shape, consistent with the purely inertial prediction~(\ref{eq:radius_dependence}). The experimental observations reported Fig.~\ref{fig:jet_profile} therefore demonstrate not only the shape similarity of the evolving jet (square-root of time spreading of the radius and hyperbolic shape), but also the validity of the assumptions made for obtaining~(\ref{eq:relaxation_wave}) and~(\ref{eq:radius_dependence}). Interestingly \citet{Stuhlman1932}, performing an \textit{a posteriori} analysis of prior experimental photographs obtained by Worthington, also evidenced an hyperbolic shape for jets following a drop impact on a liquid surface.

\noindent Using the same natural scales for velocity $\sqrt{gH}$ and length $L$ as previously, we can non-dimensionalize the prefactor $C$ occurring in the expression for the radius~(\ref{eq:radius_dependence}). Without loss of generality, the resulting dimensionless number $C/C_\text{dim}$ (noting $C_\text{dim}~= (gH)^{1/4} L^{3/2}$) only depends on the cavity aspect ratio $H/L$. This dependence is tested experimentally Fig.~\ref{fig:jet_profile}d and it is found that $C/C_\text{dim}$ is insensitive to the aspect ratio. From the experimental data we therefore conclude that the prefactor $C$ is approximately equal to $0.09 \,(gH)^{1/4} L^{3/2}$.

\subsection{Simple energy balance model}

\noindent Insights into the particular dependence of the jet velocity with the initial geometry of the cavity can be gained using first principles. First noting that the essence of jet formation is purely inviscid, we disregard in the following viscous effects (this hypothesis amounts to consider $\beta\nu \ll \alpha\sqrt{gH}H$ in Eq~(\ref{eq:dimensional_scaling_law})). The initial excess gravity potential energy associated with the initial hollow is:
\begin{equation}
U \sim \rho g L^2 H^2.
\label{eq:potential_energy}
\end{equation}
Note that we take no notice of the cavity shape-dependent prefactor, nor of the kinetic energy of the remnant motions in this elementary model as we are only interested in exponent determination.
We search in the following to express the full conversion of this initial energy budget into jet kinetic energy. Recalling that gravity has a trivial effect on the jet dynamics (see \S 2) we disregard such effects in the following for simplicity sake. The total kinetic energy of the flow can be decomposed into \textit{i)} the jet kinetic energy and \textit{ii)} the energy of the (typically dipolar) feeding flow pattern . As this last part decays with time, the asymptotic kinetic energy $T$ of the jet is the total energy of the flow. This quantity strongly depends not only on the velocity field within the jet, but also on the jet shape determined earlier. $T$ therefore reads:
\begin{equation}
T=\int_{z=0}^{V_{\text{tip}} t}{\frac{1}{2} \rho u^2(z) \pi r_{\text{jet}}^2(z)  dz} \sim  \rho C^2 V_{\text{tip}} \sim \rho \sqrt{gH} L^3 V_{\text{tip}}.
\label{eq:kinetic_energy}
\end{equation}
Expressing the full conversion of initial potential energy $U$ into jet kinetic energy $T$ finally gives:
\begin{equation}
V_{\text{tip}} \sim \sqrt{gH} \frac{H}{L}.
\end{equation}
Conversely in the tube filling experiment, the initial potential energy has the same expression (\ref{eq:potential_energy}) and the kinetic energy of the erupting liquid column is $T \sim \rho \text{V}_\text{tip}^2 L^2 H$, leading to $\text{V}_\text{tip} \sim \sqrt{g H}$, as observed in experiments.
As a conclusion we put forward that the key difference between these experiments is the flow focusing mechanism resulting from the cavity collapse, whose action is embodied in the jet shape evolution~(\ref{eq:radius_dependence}) and appearing in the expression for the kinetic energy.

\begin{acknowledgments}
L'Agence Nationale de la Recherche through its Grant "DEFORMATION" 
ANR-09-JCJC-0022-01, the \'Emergence(s) program of the Ville de Paris and the Direction G\'en\'erale de l'Armement (DGA) are 
acknowledged for their financial support. 
\end{acknowledgments}

% Bibliography
\bibliographystyle{jfm}
\bibliography{BiblioJetHollow}

\end{document}